\documentclass[11pt,twoside]{article}

\usepackage{asp2006}
\usepackage{epsf}
\usepackage{psfig}
\usepackage{lscape}
\usepackage{amssymb}
\usepackage{graphicx}

\begin{document}

\title{Exploring weak magnetic fields with LOFAR and SKA}   %%% Fill in title

\author{Tigran G.~Arshakian \& Rainer Beck}   %%% Fill in author names
\affil{Max-Planck-Institut f\"ur Radioastronomie, Auf dem H\"ugel 69,
53121 Bonn, Germany}    %%% Fill in author affiliations

\begin{abstract} %%% Abstract to run on from here.
Regular magnetic field structures can be derived from the Faraday rotation measures (RM) of polarized background sources observable at 1.4 GHz 
with the SKA. At lower
frequencies ($<250$\,MHz) polarimetry of radio sources with the
Low Frequency Array (LOFAR) will allow the investigation of
extremely small RM, to detect and map weak regular fields in halos
and outer parts of spiral galaxies, and in the interstellar and
intergalactic medium. Very little is known yet about the
number density of polarized sources at low frequencies.
Observed distributions of polarized sources at 350~MHz and 1.4~GHz and
perspectives to detect weak magnetic fields with LOFAR are
presented. Test observations of polarized radio sources with the
Westerbork Synthesis Radio Telescope (WSRT) and the Giant Metrewave
Radio Telescope (GMRT) are discussed.
\end{abstract}

%%% MAIN BODY OF TEXT GOES HERE. CONSULT "INSTRUCTIONS FOR AUTHORS USING
%%% LATEX2E MARKUP", SECTIONS 2.3-2.6 FOR HELP WITH EQUATIONS, FIGURES,
%%% AND TABLES.

\section*{Introduction}   %%% Top level section head (remove "%" symbol)

Most of what we know about galactic magnetic fields comes through
the detection of radio waves. {\em Synchrotron emission} is related
to the total field strength in the sky plane, while its polarization
yields the orientation of the regular field in the sky plane and
also gives the field's degree of ordering. Incorporating {\em
Faraday rotation} provides information on the strength and direction
of the coherent field component along the line of sight.

Low frequency radio emission is purely nonthermal in nearby galaxies
and in intergalactic (IGM) and intracluster magnetic media. Radio
emission detected with present day radio telescopes allows detailed
studies of a few dozens of nearby galaxies \citep{beck05}. With the
high sensitivity of new generation telescopes such as the Low
Frequency Array (LOFAR) and the Square Kilometre Array (SKA), one
can observe very weak low-frequency ($\la 250$ MHz) radio emission
and hence measure weak magnetic fields in the outer parts of
galaxies and the IGM.

{\em The polarimetric capabilities} of LOFAR and the SKA are crucial
for our understanding of magnetic fields. The required
specifications are high polarization purity and multichannel
spectro-polarimetry. The former will allow the detection of objects
with relatively low degrees of linear polarization, while the latter
will enable accurate measurements of Faraday rotation measures
(RMs), intrinsic polarization angles and Zeeman splitting.
The method of {\em RM Synthesis}, based on multichannel
spectro-polarimetry, transforms the spectral data cube into a data
cube of Faraday depth \citep{brentjens05}. A range of different RM
values can be observed and RM components from distinct regions along
the line of sight can be separated.

%Advantages of the low-frequency radio astronomy.
%LF emission is purely nonthermal in nearby galaxies, IGM and ISM.
%Radio synchrotron emission is a measure of the strength of the
%total magnetic field ($B_{\rm tot}$):
%
%Allows detailed studies for few dozens of nearby galaxies … .
%
%Linear polarization – degree of ordering of the magnetic field:
%   Fully ordered field can polarize the signal up to 75%.

%Small Rotation Measures (RM ~ ? ne B|| dr) can be measured (?RM ~ ?-2) ? weak magnetic fields.

%RM Synthesis \citep{brentjens05} –  separate RM components from
%regions along the LOS (from multichannel spectro-polarimetry).

\section*{Synchrotron radio emission from weak magnetic fields}

\begin{figure}
\centering
\includegraphics[angle=-90,width=10cm]{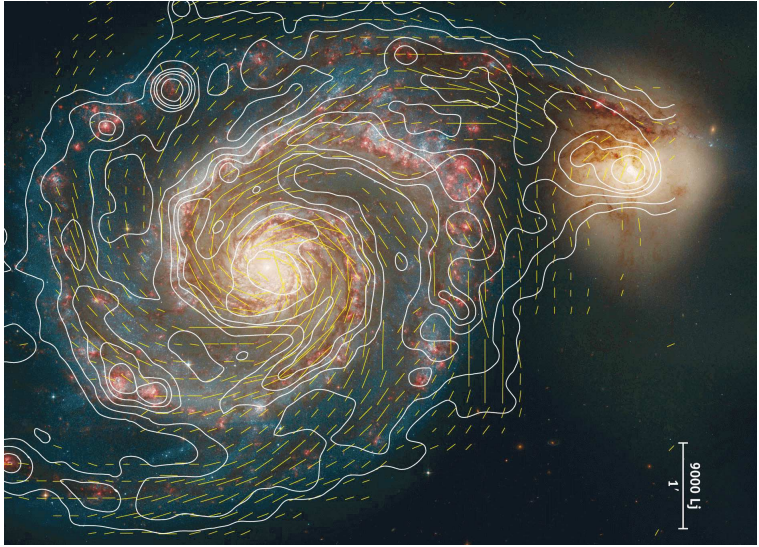}
\caption{HST image of the spiral galaxy M~51, overlaid by contours
of the intensity of total radio emission at 6.2~cm wavelength and
B--vectors, combined from data from the VLA and Effelsberg 100~m
telescopes and smoothed to 15'' resolution (Fletcher et al., in
prep.) (Graphics: Sterne und Weltraum. Copyright: MPIfR Bonn and
Hubble Heritage Team).} \label{fig:m51}
\end{figure}

Synchrotron radio emission at low radio frequencies is radiated by
cosmic-ray electrons which are accelerated by supernova shock fronts
in star-forming regions. The diffusion length of cosmic-ray
electrons is longer at low frequencies because energy losses of
electrons are smaller. The synchrotron lifetime of electrons
emitting in the LOFAR bands ($30-240\,$MHz) is $(2-5)\times10^{8}$
yr for a magnetic field strength of $B=5\,\mu$G. Cosmic-ray
electrons radiating at 50\,MHz can propagate to a large distance
(up to 150\,kpc) from the place of origin and reach outer
regions of the disk and halo of galaxies \citep{beck08}. Weak
magnetic fields in outer galactic regions, in the halos of clusters
of galaxies, and in the intergalactic space can be traced by
polarized synchrotron emission if ordered magnetic fields exist. At present, regular magnetic fields are observed in nearby
spiral galaxies (Fig.~\ref{fig:m51}) and in cluster relics
\citep[e.g.][]{govoni05}. The large scalelength and/or scaleheight of magnetic
fields in the disks and halos of galaxies suggests that weak regular
fields may also exist in the outer regions \citep{beck05}. More
sensitive radio telescopes, such as LOFAR and the low-frequency SKA
array, are needed to detect a weak polarized signal from outer parts
of galaxies, clusters and the IGM.

%\subsection{}   %%% Second level section head (remove "%" symbol)

%\subsubsection{}   %%% Lowest level section head (remove "%" symbol)

\section*{Faraday rotation measure as a tool to study weak magnetic fields}
%%% Unnumbered top level section head (remove "%" symbol)

Another potential tool for studying weak magnetic fields in galaxies
is the Faraday rotation measure (RM) of background polarized sources
shining through the foreground galaxy. RM in galaxies are generated
by regular fields of the galaxy plus its ionized gas, both of which
extend to large galactic radii (Fig.~\ref{fig:m51}). RM towards
polarized background sources can trace regular magnetic fields in
these galaxies out to even larger distances. However, with the
sensitivity of present-day radio telescopes the number density of
polarized background sources is only a few sources per square degree
solid angle, so that only M~31 and the Large and Small Magellanic
Clouds could be investigated in this way so far \citep{han98,gaensler05,mao08}.

With the high sensitivity of the SKA at 1.4\,GHz, we will observe
the polarized intensity and RM for a huge number of faint radio
sources, thus providing a high-density background of polarized point
sources. This opens the possibility to study in detail large-scale
patterns of magnetic fields in galaxies. By measuring RM at frequencies
around 1~GHz with the SKA, simple field structures can be recognized
in galaxies up to about 100\,Mpc distance and will allow to test
dynamo against primordial or other models of field origin for $\le
60000$ disk galaxies \citep{stepanov08}. For this, a source sensitivity of $0.015$~$\mu$Jy of integrated flux density is
needed, which can be achieved within 100~h observation time with the
SKA. On the other hand, the reconstruction of magnetic field
structures of strongly inclined spiral galaxies would require a much
higher sensitivity of the SKA \citep{stepanov08}. The field structures 
of $\approx 60$
galaxies to about 10~Mpc distance can be recognized with $\ga 1000$
background polarized sources which would require tens to hundred
hours of integration time. 
%Recognition or reconstruction of regular
%field structures from RM data of polarized background sources is a
%powerful tool for future radio telescopes such as the SKA

As the RM errors are smaller at larger wavelengths ($\Delta RM
\propto \lambda^{-2}$ for constant relative bandwidth) the
low-frequency SKA array and low-frequency precursor telescopes like
LOFAR will be ideal to study the weak magnetic fields in galaxies
and intergalactic space -- provided background sources are still
significantly polarized at low frequencies.

\subsection*{Polarization properties of radio galaxies at 150\,MHz}

The main population of galaxies contributing to the polarized
background sky at low radio frequencies are FR\,I and FR\,II radio
galaxies. \cite{schoenmakers00} observed the double-double giant
radio galaxy 1834+62 at frequencies from 600\,MHz to 8.4\,GHz. They
found strong polarization in all four radio lobes (19\% - 22\%) at
1.4\,GHz. The rotation measures towards four radio lobes are
similar, within $\sim (3\pm0.4)$ rad m$^{-2}$ of $57$ rad m$^{-2}$.
This difference is hard to explain by RM produced by cocoons
surrounding the lobes, and it is most likely originating in the
diffuse medium of the Galaxy and/or intergalactic space or
intergalactic filaments.

To study the polarization properties of this source at lower
frequencies we conducted observations with the WSRT
at 150\,MHz and 350\,MHz (joint project with Ger de Bruyn) with 
resolutions of 130" and 55", respectively.
Preliminary results show that the inner lobes are depolarized at
350\,MHz, while the outer lobes still show 13\% polarization. At
150\,MHz the outer lobes are more depolarized, to $< 5\%$.
Fractional polarization of outer lobes at 150\,MHz is $\ga2$
times smaller than at 350\,MHz and $\ga4$ times at 1.4\,GHz suggesting that
beam and/or depth depolarization is stronger at 150\,MHz. Therefore, 
the number density of polarized sources will be low, at least for
large beams.

The polarized sky was observed within the primary beam of the GMRT (using the experimental software backend) near 150\,MHz at 20" resolution \citep{pen08}.
They showed that the level of polarized sky brightness at 150\,MHz
is well below that expected from extrapolation at higher
frequencies, and that fractional polarization of extragalactic point
sources is $< 1\%$, indicating strong depolarization.

\subsection*{Density of polarized background sources at 350\,MHz}

LOFAR can in principle measure very small Faraday rotation measures
(below $\lesssim 1$ rad m$^{-2}$) and hence very weak magnetic
fields and/or small electron densities in the outer disks of 
galaxies ($RM<10$ rad m$^{-2}$), galaxy
halos, cluster relics and in intergalactic filaments ($RM \sim 1$
rad m$^{-2}$), and, possibly, also in the general intergalactic
medium($RM \lesssim 0.1$ rad m$^{-2}$).

\begin{figure}
\centering
\includegraphics[width=10cm]{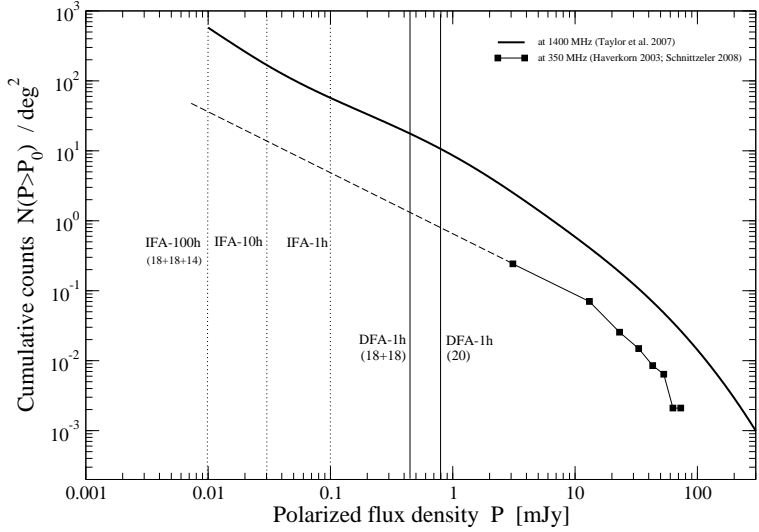}
%\resizebox{\hsize}{!}{\includegraphics{N-P_350MHz_12.eps}}
  \caption {Observed integral number counts of polarized background
  point sources at 350\,MHz (black squares) and 1.4\,GHz
  \citep[][solid line]{taylor07}. The dashed line is the extrapolation
  of number counts at 350\,MHz down to flux density level of 0.01\,mJy.
  The two vertical solid lines represent the $5\sigma$ detection limits with
  the LOFAR Dutch Full Array within 1\,h observation time for 20 (\emph{right})
  and 36 LOFAR stations (\emph{left}).
  The three vertical dotted lines are the $5\sigma$ detection limits with
  the Dutch Full Array and International Full Array (36+14 stations) in the ``high-band''
  (110--240\,MHz) within 1\,h (\emph{right}), 10\,h (\emph{middle}) and
  100\,h (\emph{left}) observation time.}
  \label{fig:ncp}
\end{figure}

%LOFAR and the low-frequency SKA array can detect smaller RM values
%and recognize weak galactic and intergalactic magnetic fields if
%background sources are still polarized at low frequencies
%($<350$\,MHz). 
As the flux density decrease with frequency (for
steep-spectrum sources), one should expect more polarized background
sources to be detected at lower frequencies. On the other hand, the
depolarization is stronger at low frequencies which weakens the
polarized signal. Deep low-frequency polarization surveys are needed
to determine the cumulative source counts.

Because of the lack of polarization surveys at low frequencies
($<600$\,MHz), we used several fields observed with the WSRT at
350\,MHz
\citep{haverkorn03a,haverkorn03b,schnitzeler08,schnitzeler09} to
compile a sample of 104 polarized extragalactic sources. The number
counts of polarized sources at 350\,MHz (Fig.~\ref{fig:ncp}; black
squares) is one order of magnitude lower than that at 1400\,MHz
(Fig.~\ref{fig:ncp}; solid line). This suggests that at low
frequencies depolarization is larger than the spectral increase and
reduces the polarized flux, and hence the observable number of polarized
background sources. Extrapolation of the number counts down to
0.01\,mJy (the dashed line in Fig.~\ref{fig:ncp}) shows that with a
sensitivity of 0.5\,mJy the high-band LOFAR will detect $\la 1$
polarized source per 1 deg$^2$ and $\approx 5$ sources in 1 hours
integration time with the Dutch and International Full Array (DFA
and IFA) configurations, respectively. From 10 to 30 polarized
sources will be detected with IFA in 10 to 100 hours of integration
time.

RM mapping with LOFAR is possible for nearby sources with large
angular sizes covering a sky area of several square degrees. The
high-band antennae (110-240\,MHz) are preferable because calibration is easier
while detection of polarization in the
low-band (30-80\,MHz) will be much harder. The Coma Cluster of
galaxies covers an area of about 50\,deg$^2$, and LOFAR (high-band)
will detect $\approx 500$ background polarized sources towards the
cluster in 10 hours of integration time, thus allowing detailed RM
mapping and studying of the structure of the large-scale magnetic
fields. About 30 polarized sources can be detected behind a
filamentary structure ($\approx 3$ deg$^2$), found in the
central part of the Coma cluster at 408\,MHz \citep{kronberg07}, and
about 80 polarized sources behind the nearest galaxy, M\,31, with
100 hours of integration time. The latter number of background
sources is four times larger than that detected with the VLA at
1.4\,GHz \citep{han98}, thus allowing the detailed structure of the
field to be investigated.

\section*{Summary}

Extremely small RM values can be measured in the LOFAR high-band
(110-240\,MHz), thus providing a powerful tool to measure weak
magnetic fields. Recognition of simple magnetic field structures of
foreground objects is realistically possible with $\ga 10$ polarized background
sources per object. Depolarization is stronger at lower radio frequencies which
leads to a strongly reduced number density of polarized point
sources. Therefore, recognition and detection of weak magnetic field
structures is only possible for nearby sources with large angular
sizes covering a sky area of several square degrees (nearby
galaxies, clusters and filaments), while distant objects (or objects
with small angular sizes) require much observation time ($\ga 100$\,h).
For objects with small angular sizes, the diffuse polarized emission
from the objects themselves has to be detected to measure their RM.

%\acknowledgements
%%% Text of acknowledgements runs on after this command.

%%% THE BIBLIOGRAPHY
%%%
%%% CONSULT SECTION 3 OF "INSTRUCTIONS FOR AUTHORS" FOR HOW TO USE NATBIB.
%%% AUTHORS ARE ENCOURAGED TO USE EITHER THE "THEBIBLIOGRAPY" ENVIRONMENT
%%% BY UNCOMMENTING (DELETING THE "%" SYMBOL) THE COMMANDS BELOW, OR BY
%%% USING THE BIBTEX ENVIRONMENT. TO FIND OUT WHICH IS APPLICABLE TO YOUR
%%% CONTRIBUTION, CONSULT THE VOLUME EDITORS FOR YOUR PROCEEDINGS.
%%%

\end{document}